# Omitting Uncertainty of Thermal Conductivity Measurement in Liquids and Nanofluids by Steady-state Circular Parallel-plates

A. Hosseinpour[1], N. Piroozfam[2], N. E. Kandjani[2], E. Esmaeilzadeh[1, 2] *

1) Islamic Azad University, Tabriz Branch, Mechanical Engineering Department, Tabriz, Iran

2) University of Tabriz, Mechanical Engineering Department, Tabriz, Iran

*) Corresponding author (esmzadeh@tabrizu.ac.ir)

**Abstract**

In this paper, thermal conductivity measurement of liquids and nanofluids is numerically investigated by the steady state method. In which, it is required to omit the natural convection effects on the inaccuracy of measurement for heat lost quantification. One of the well-known steady-state apparatus is circular parallel-plate where the distance between circular plates is filled with liquids or nanofluids. So, it has to be designed accurately to neglect the effects of free convection. The numerical simulation is a convenient method which can omit the mentioned effects. Water, Ethylene and Ethylene Glycol are considered as base liquids and Al2O3 as a Nano-fluid in this work. Thermal boundary conditions, the distance between two parallel discs, physical and thermal quantities of fluids, size and volume fraction of nano-particles have extensive effects on the measurement process. The main result of the simulation indicates that for all of the cases, the distance between two parallel discs has to get less than 2 mm for accurate measurement of thermal conductivity in the parallel circular plate, wherein, uncertainty in the presence of the natural convection can be easily omitted.

## 1- Introduction

Heat transfer increment has always been one of the hardest challenges of the thermal engineers. Due to the advent of the new technologies, transferring heat needs to be at higher rate and efficiency from a smaller area or across a lower temperature difference. Therefore, various techniques on the heat transfer augmentation have been suggested. The thermal conductivity's order of magnitude in most of the heat transfer fluids is lower than the one in solid metals. The addition of micro-sized solid metal or metal oxide particles to the base fluids demonstrates a rise in the thermal conductivity of resultant fluids [1]. Nevertheless, the solid particles with their sizes mentioned before have many disadvantages in the transport applications. Nanofluids can be potentially applied to the fields such as microelectronics, transportation, manufacturing, and medical; regarding their well-improved thermal properties. They are imperative in energy utilization and sustainable technology development because of not having any defective effects.

 The volume addition of only a small percent of solids produces a dramatic increase in thermal conductivity [2-5]. The works that have been done in the area of nanofluids up to now are assigned to synthesis, characterization, and some applications in convective heat transfer and boiling. Experiments show that thermal conductivity of nanofluids depends on a large number of parameters namely, the chemical composition of the solid particle and the base fluid, particle concentration, particle shape and size, surfactants, etc. Techniques in measuring nanofluids' effective thermal conductivity are mostly used as the transient hot-wire [3], steady-state [4], cylindrical cell [5], temperature oscillation [6], and 3-omega [7].

The steady-state method from the above can be dominant if it is aptly designed for the following reasons:

- Low cost of the measurement device,

- Simple fluid and nanofluid sampling for measuring thermal conductivity,

- Shorter measurement time through limited temperature data transfer,

- Control of the temperature distribution in the radial direction.

Due to lack of the proper designs in previous works, uncertainty analyses were conspicuously different from the results of the other methods. Therefore, most of the researchers and technical experts are not sufficiently confident in the use of these methods. However, if a study platform is appropriately designed and nanofluid sampling is suitably conducted, measurement results can be trusted. In most of the studies, measurement methods characteristics are mentioned without considering the measurement method accuracy or uncertainty analysis. Therefore, most of the presented results are not almost convergent, and only the norms of behavior are discussed.

According to the miniature device proposed in the present work for reducing restrictions and factors that increase errors in the measurement, achieving the correct result is by numerical calculation. The main limitation of this way is the simultaneous presence of conduction and natural convection heat transfer. The second one needs to be removed to achieve the correct result to consider the fluid and nano-fluid mass sample as a semi-rigid body. This can only be gained through applying fixed heat flux on the lower surface and constant temperature at the upper surface with several temperature measurements to accurately calculate the thermal conductivity of the fluid and nanofluid. Taking into consideration that in the design and selection of the survey, the distance between two disks is assumed much smaller than their radii so that the radial temperature gradient can be ignored. Several studies have been conducted on the importance of thermal conductivity of nano-fluid in the heat transfer process, some of which are discussed in the following.

Based on the steady-state heat transfer, various designs of the test cells can be used to measure the thermal conductivity of fluids. To facilitate heat transfer, mainly two parallel plates or cylindrical cells are preferred.

The apparatus for the steady-state parallel-plate method can be constructed by the design of Challoner and Powell [8]. They have used it for measuring the thermal conductivity of alumina and copper oxide based nanofluids. Masuda et al. [9] studied the thermophysical properties of the metallic oxide particles (Al2O3 and TiO2) dispersed in water. The transient hot-wire method was used for measuring the thermal conductivity of nanofluids. They reported that the thermal conductivity of nanofluids was significantly larger than the base liquid. Murshed et al. [10] measured the thermal conductivity of TiO2 nanoparticles in rod shapes and spherical shapes dispersed in deionized water. Bruggeman [11], and Wasp [12] used the transient hot-wire apparatus to measure the thermal conductivity of the nanofluids which was compared later to the theoretical prediction models of Hamilton and Crosser [13]. The results showed that the thermal conductivity of nanofluids increases remarkably with increment of nanoparticles' volume fraction. Furthermore, the size and shape of particles influence the thermal conductivity enhancement of nanofluids. Das et al. [14] reported the thermal conductivity of Al2O3 and CuO nanoparticles suspended in water as a function of temperature. The temperature oscillation technique was used in their study for measuring the thermal conductivity of nanofluids at different temperatures ranging from 21°C to 51°C. The results depicted that thermal conductivity had been raised with nanofluids temperature augmentation as well as particle concentrations. Eastman et al. [15] reported the thermal conductivity of nanofluids containing Al2O3, CuO, and Cu nanoparticles with two different base fluids: water and HE-200 oil. 60% improvement of the effective thermal conductivity was achieved as compared to the corresponding base fluids for only 5% of the volumetric concentration of nanoparticles. Lee et al. [16] suspended CuO and Al2O3 using two different base fluids; water and ethylene glycol (EG) and obtained four combinations of nanofluids. Their results proved that nanofluids have significantly higher thermal conductivities than the base fluid. They have more than 20% at 4% of the volumetric concentration of CuO nanoparticles with EG as the base fluid. In the low volume concentration ranges, the effective thermal conductivity increases almost linearly with volumetric concentration.

The key objective of this study is to investigate the free convective heat transfer in a closed environment under various thermal conditions. Regarding this, a lot of study through natural convection heat transfer in nanofluids has been done, but, these studies have an aspect of the review and their objectives are not related to the measurement. Sheikhzadeh et al. [17] numerically studied the natural convection of copper oxide-water nanofluid in a square cavity with hot and cold sources on vertical walls and reported that heat transfer raises with increasing volume fraction of nanoparticles. Ternik et al. [18] examined the steady natural convection in a cubic cavity filled with gold-water nanofluid using the finite volume method. They observed that the start of convection is delayed by adding nanoparticles, and heat transfer rate can be improved based on the volume fraction of nanoparticles and Rayleigh number.
Therefore, in this study, variables governing the problem were entered to investigate the behavior of fluids and nanofluids in the application of thermal conductivity. The measurement method in the steady state of two parallel discs that are at different distances, presents a new view in using this simple and low-cost method.

## 2. Numerical Simulation

### 2.1. Computational Domain and Boundary Conditions

The cylinder is obtained from two parallel disks in our geometry. Side view of the geometry is modeled in the form of two parallel discs with different cavity thicknesses. The dimensions of the part are the variables of the problem. By changing them, a convenient dimension where heat transfer is just by conduction is obtained. In the first step, the fluid sample was placed between two discs. Thermal conductivity and suitable dimensions for the heat transfer by conduction was found. Second, the base fluid was changed to ethylene glycol and in this case, the expression was evaluated. Finally, to improve the heat transfer, the addition of nanoparticles to the base fluid was used. Aluminum oxide ($Al_2O_3$) nanoparticles have been used and their thermal conductivity was found. Various nano-particle volume fractions from 0.2-7% were studied. According to the dimensions and geometric conditions, the temperature distribution is two-dimensional and we have:

$$T_f = T_f(r \cdot y)$$

The present study seeks to eliminate the effects of the following:

$$\frac{\partial T}{\partial r} \cong 0 \text{ and } Ra = Pr \times Gr < 10^3$$

In order to omit the effects of natural convection, fluid and Nano fluid can be assumed to be semi-rigid body under extreme conditions. So, the basic variables of the present work, are two geometric quantities r and d. In the original model, the disk radius of 25 mm is considered (Figure 1). By calculating K (W/m. k), it was observed that the diameter of the disc does not have a huge impact on thermal conductivity. And, to reduce the calculation, a model with a radius of 2.5 mm is considered. The side walls of the cavity are defined by isolating boundary condition. However, the heat transfer from the other walls is possible. The upper boundary of cylindrical cavity is studied under constant temperature conditions and it is taken 280.15 k. The lower border of the cylindrical cavity has been studied under constant heat flux and chosen 3000 W.

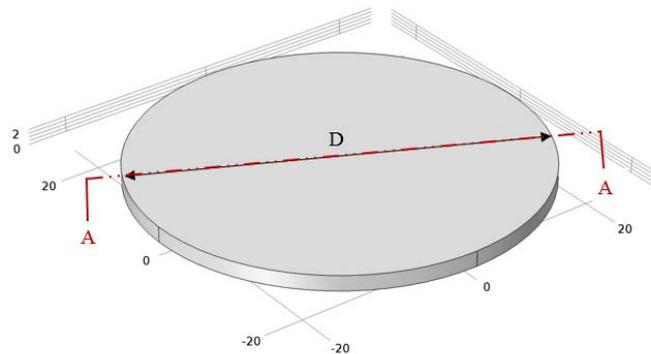

a) 3D view of the model with a diameter of 50 mm and a height of 2 mm

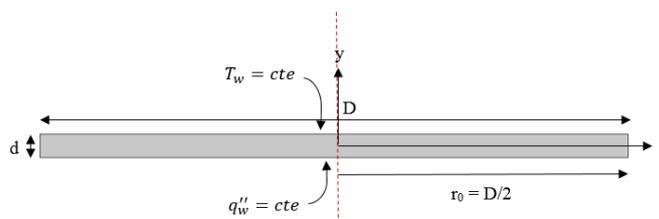

b) A_A point of view of the 3D mode

**Figure 1**: Schematic figure of the present model, a) 3D view, b) cutting A_A

## 2.2. Grid Generation

The computational domain should be discrete for the numerical solution. Among triangle, square and hexagonal grids; triangular grid gives more accurate result. Hence, it is selected. To check the needed cell number to get the exact answer, the coefficient of thermal conductivity is utilized. Four sizes of this type of grid in cavity with dimensions of 2 × 5 mm with the water base fluid were surveyed and finally, according to the results, 31718 cell number is selected.

## 2.3. CFD Modeling Strategy

In this paper, non-isothermal flow has been chosen including two subsections, laminar flow and heat transfer. Due to the physical conditions of the model, the steady state solution is the best choice. The finite element method is utilized for numerical calculation. This method is based on the elimination of differential equations or simplifying them to ordinary differential equations solved with the numerical method. The calculations for each cavity will continue until the remaining equations of continuity, momentum and energy reach to $10^{-6}$.

## 2.4. Governing Equations

In this section, noting the lower volume fraction of particles, the single-phase method is used to solve the nanofluid flow. In the single-phase conservation equations, homogenous model similar to conventional fluid is considered. In this way, the slip between the base fluid and nanoparticles will not be considered. And, the presence of particles can be seen in the extended properties. The conservation equations are presented in the following form:

Continuity:

$$\nabla \cdot (\rho_{\text{eff}} \vec{V}) = 0 \qquad (1)$$

Momentum:

$$\nabla \cdot (\rho_{eff} \vec{V}\vec{V}) = -\nabla p + \nabla \cdot (\mu_{eff} \nabla \vec{V}) + \rho_{eff} \vec{g} \beta (T - T_i) \qquad (2)$$

Energy:

$$\nabla \cdot (\rho_{eff} c_{peff} \vec{V} T) = \nabla \cdot (K_{eff} \nabla T) \qquad (3)$$

Thermal and physical properties are based in their effective values and are shown with subscript *eff* in equations.

For density, specific heat, dynamic viscosity and thermal conductivity of nanofluids the following relations are used [19]:

$$\rho_{nf} = (1 - \varepsilon_p)\rho_{bf} + \varepsilon_p \rho_p \qquad (4)$$

$$(c_p)_{nf} = (1 - \varepsilon_p)(c_p)_{bf} + \varepsilon_p(c_p)_p \tag{5}$$

$$\begin{cases} \mu_{nf} = \mu_{bf}(123\varepsilon_p^2 + 7.3\varepsilon_p + 1) & \text{Water -Al}_2\text{O}_3 \\ \mu_{nf} = \mu_{bf}(306\varepsilon_p^2 + 0.19\varepsilon_p + 1) & \text{EG-Al}_2\text{O}_3 \end{cases} \tag{6}$$

$$\begin{cases} K_{nf} = K_{bf}(4.97\varepsilon_p^2 + 2.72\varepsilon_p + 1) & \text{Water-Al}_2\text{O}_3 \\ K_{nf} = K_{bf}(28.905\varepsilon_p^2 + 2.827\varepsilon_p + 1) & \text{EG-Al}_2\text{O}_3 \end{cases} \tag{7}$$

In all the above relationships, ε is the volume fraction of nanoparticles. Thermophysical properties stated above are functions of temperature as well.

For comparing analytical results with the results of our solutions, the following equation is used:

$$Q = k \frac{\Delta T}{\Delta x} \tag{8}$$

In this regard, Δx, is cylindrical cavity thickness. To determine the scope of heat conduction and natural convection, the Ra equation will be used. For heat transfer by conduction, the Ra number should be less than $10^3$.

$$Ra = Gr \times Pr \tag{9}$$

The properties of the base fluid and the aluminum oxide nanoparticles are given in Table 1.

Table 1: Properties of nanoparticles and base fluid [20,21]

| Physical properties | Water | EG | Al$_2$O$_3$ |
|---|---|---|---|
| $C_p (J/kgK)$ | 4179 | 2415 | **765** |
| $\rho (kg/m^3)$ | 997.1 | 1114 | **3970** |
| $K (W/mK)$ | 0.613 | 0.252 | **25** |
| $\beta \times 10^{-5} (1/K)$ | 21 | 57 | **0.85** |
| $d_p (nm)$ | 0.384 | 0.561 | **47** |

Particle geometry plays an important role in thermal conductivity of the mixtures and for this reason, effective thermal conductivity of mixtures is expressed as follows [22]:

$$K_{nf} = K_{bf}\left[1 + \frac{K_p \times \varepsilon \times r_{bf}}{K_{bf} \times (1-\varepsilon) \times r_p}\right] \tag{10}$$

Where, r$_p$ is particle radius and r$_{bf}$ is half of liquid thickness.

## 3. Results and Discussion

### 3.1. Validation

To verify the results of this work, the original version of the water base fluid with the results of the Eckert [23], which has been shown in the table 2 , was compared. The results of this comparison are visible in Figure 2. As it can be seen, owing to a very small error between the data from this study and the Eckert, the accuracy of the results can be trusted.

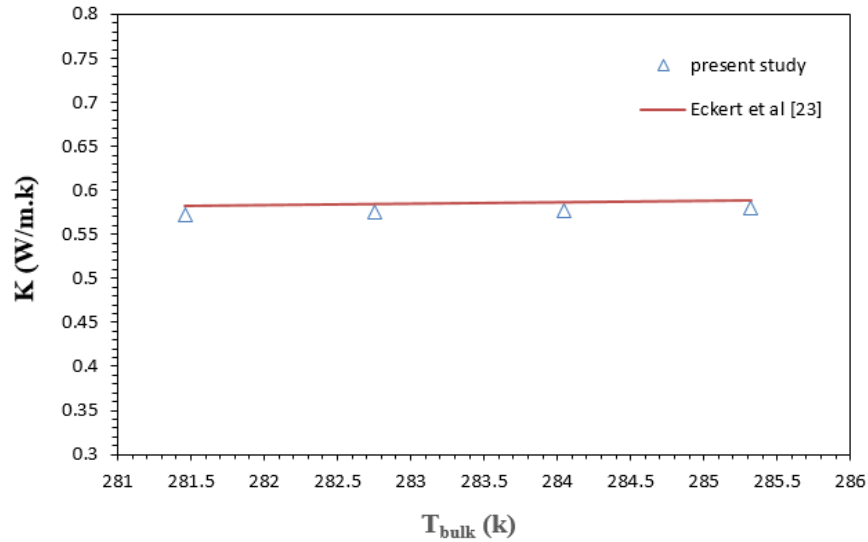

**Figure 2:** validation of water thermal conductivity by studying of Eckert et al [23]

Values for the thermal conductivity of this study and the results of the study of Eckert et al. were compared. And the comparison of the results is shown in Table 2. According to it, for the water base fluid about 1.5% error and for ethylene glycol based fluid less than 1% error was seen.

**Table 2:** Comparison of the results of current study for thermal conductivity of Ethylene Glycol with Eckert et al. [23]

|          | 292.195 | 289.22 | 286.22 | 283.2  | $T_{bulk}$ (k) |
|----------|---------|--------|--------|--------|----------------|
| **K (W/m.k)** | 0.2486 | 0.2476 | 0.2465 | 0.2455 | **Eckert et al.** |
| **K (W/m.k)** | 0.2490 | 0.2480 | 0.2470 | 0.2450 | **Current study** |

## 3.2. Results of Numerical Simulation

### 3.2.1. Base Fluids

The two fluids, water and ethylene glycol as a base fluid are selected. The different distance between two parallel discs was studied to find an optimal distance in which the heat transfer is obtained just by conduction. The amount in the previous empirical studies for an optimal thickness, 1.21 mm was introduced [24]. The experimental methods and laboratory studies have been conducted in these works. Using numerical study conducted in this paper with the utmost precision, optimum thickness of the cavity for the heat transfer by conduction is 2 mm. The water fluid streamlines can be seen in various dimensions of the cavity in Figure 3.

As it can be observed, for thickness less than 2 mm, only heat conduction has occurred. Flow lines are small cut-lines that indicate the constant fluid level and very slow motion of the flow. In dimensions of more than 2 mm, the flow lines are reversed. This means that the natural convection heat transfer occurs. The return flow lines demonstrate flow on a rotating basis, which occurred as a reason of the natural convection. Similar behavior is observed for ethylene glycol base fluid.

For the heat transfer only by conduction, value of the Rayleigh number range is smaller than $10^3$. To implement the results, for the maximum thickness of the cylindrical cavity wherein heat transfer is only by conduction, Figure 4 was drawn. Two zones are shown for certainty and uncertainty of two essential parameters of liquids and nanofluids thermal conductivity measurements by this method. In zone (**I**) it is required to limit the values of thickness and Rayleigh numbers as: y<2mm and Ra<1000 for omitting the free convection effects. For all values of greater than the previous mentioned limits, the measurements are located in non-accepted zone (**II**).

In Figure 5, isothermal lines for water base fluid is shown in various distances between two discs. As it can be seen in *a* and *b*, isothermal lines are horizontal. This means that the temperature is constant at any point of *r* axis, which represents that the heat transfer in *y* direction is only by conduction mode. By increasing of the cavity thickness, difference between high and low temperatures will be increased and isothermal lines will change shape.

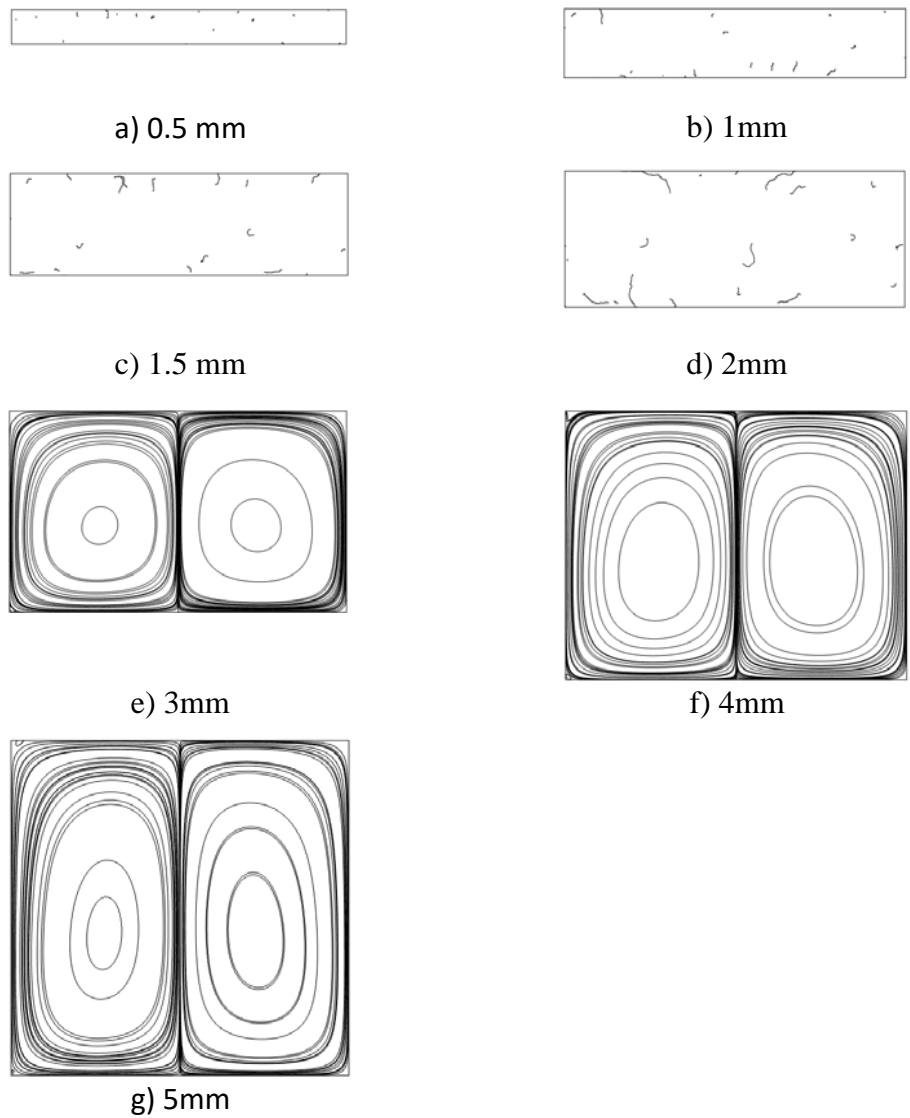

**Figure 3:** Ethylene Glycol base fluid streamlines in thickness of cavity

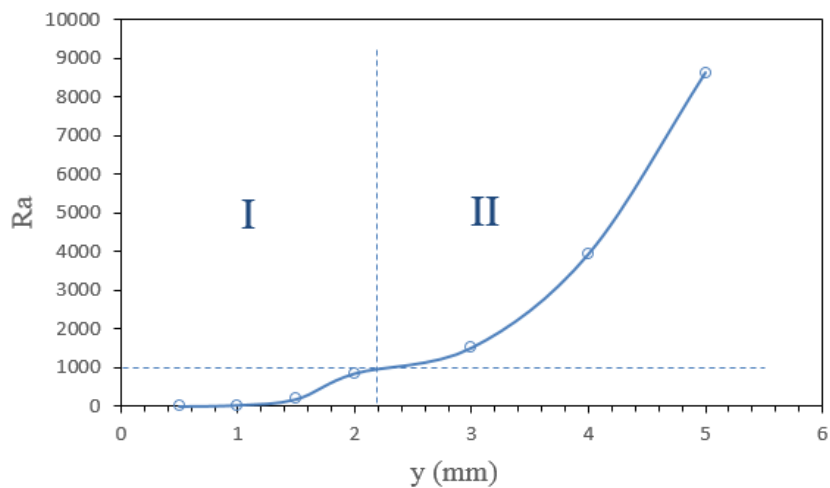

**Figure 4:** Comparison of the number of Rayleigh in different thickness of cylindrical cavity

### 3.2.2. Water-Al$_2$O$_3$ Nanofluid

Nanoparticles are added to the base fluid in order to raise the conductivity of the fluid inside the cavity. In this study, utilized nanoparticle is aluminum oxide (Al$_2$O$_3$). It is added to water base fluid with different range of volume fraction from 2-7 %.

In Figure 6, correlation between conductivity coefficient and volume fraction is obtained in four distances (noting that the conduction heat transfer only takes place). As can be seen from the diagram, by adding the nanoparticle to the base fluid, thermal conductivity shows a significant growth. Furthermore, the thermal conductivity increases with the volume fraction increment. Mentioned behavior is valid for all thicknesses.

To better understand the flow through cylindrical cavity, velocity profiles in various thicknesses of cavity for aluminum oxide–water nanofluid for 3% volume fraction are plotted in Figure 7.

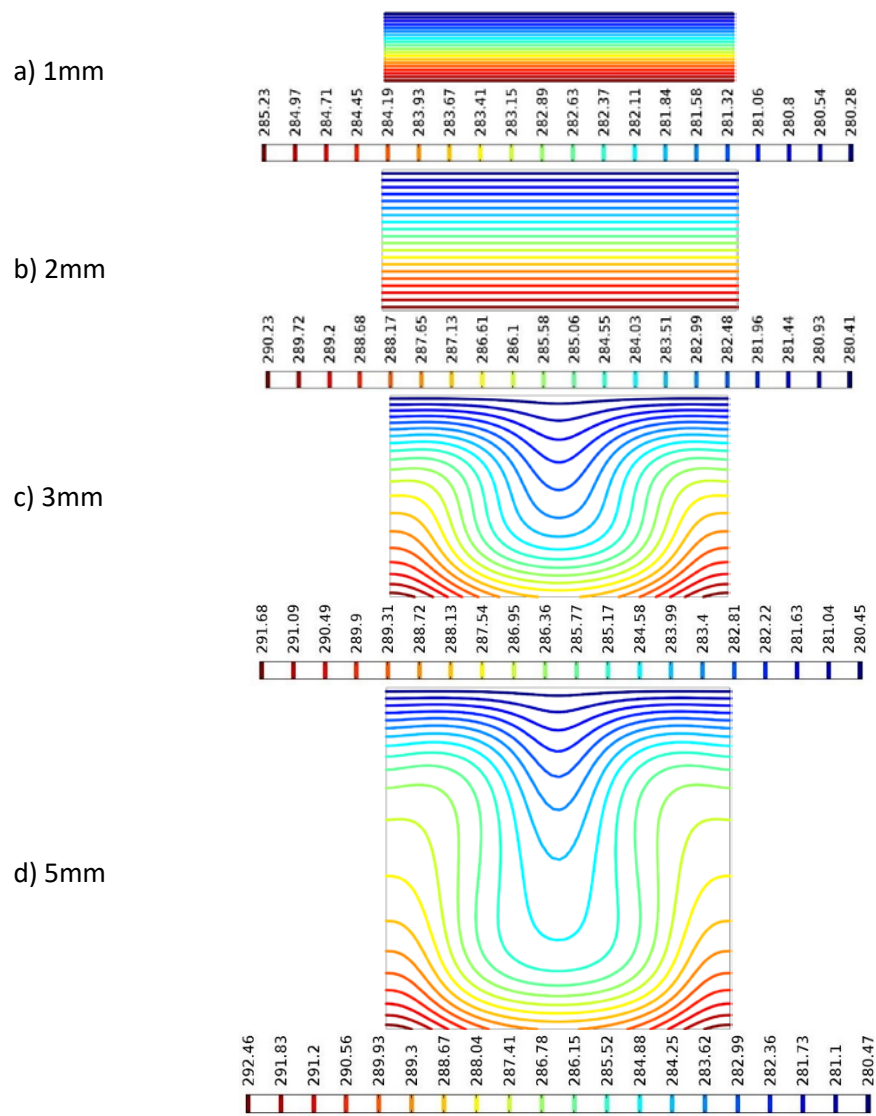

**Figure 5:** Isothermal lines in the water base fluid for different thicknesses of cavity

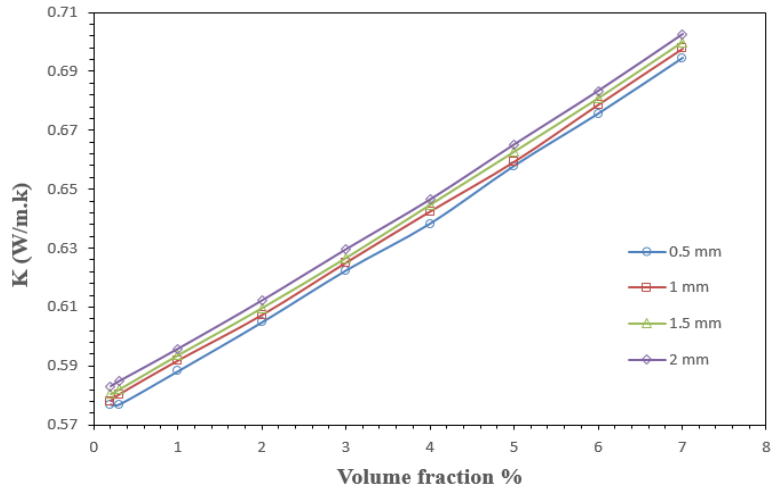

**Figure 6:** Changes in thermal conductivity of Water- Aluminum Oxide nanofluid in various thicknesses of cavity

As can be observed in velocity profiles, in thickness of less than 2 mm, fluid flow does not show any specific move. By increasing the thickness of the cylindrical cavity, rotational flow is created which represents the natural convection heat transfer.

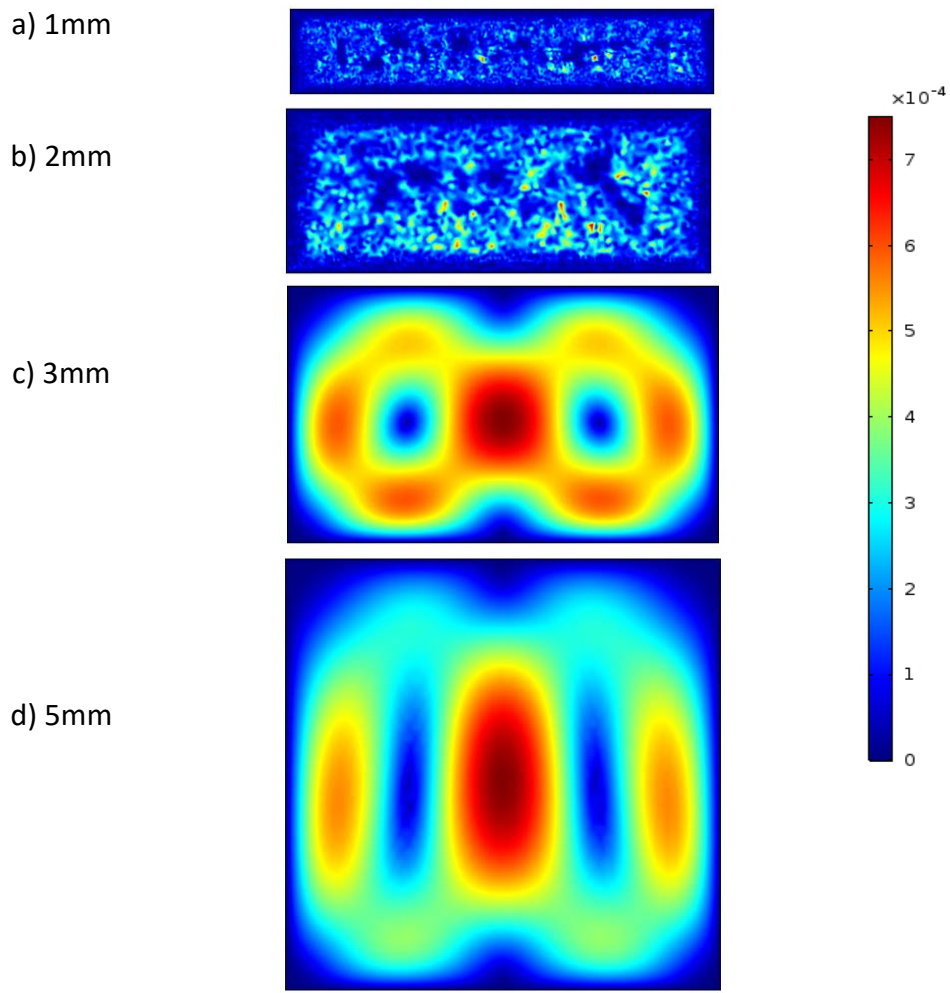

**Figure 7:** Velocity profiles for Water- Aluminum Oxide nanofluid in various thicknesses of cavity

In Figure 8, velocity changes in the thickness of 2 mm for cylindrical cavity with different volume fractions are plotted. In all cases, the values of velocity are very low which means that motion of the nanofluid is very slow. With volume fraction augmentation, the average velocity of the flow is lessened. According to the diagram, it can be understood that the maximum velocity in the cavity occurs in the lower half.

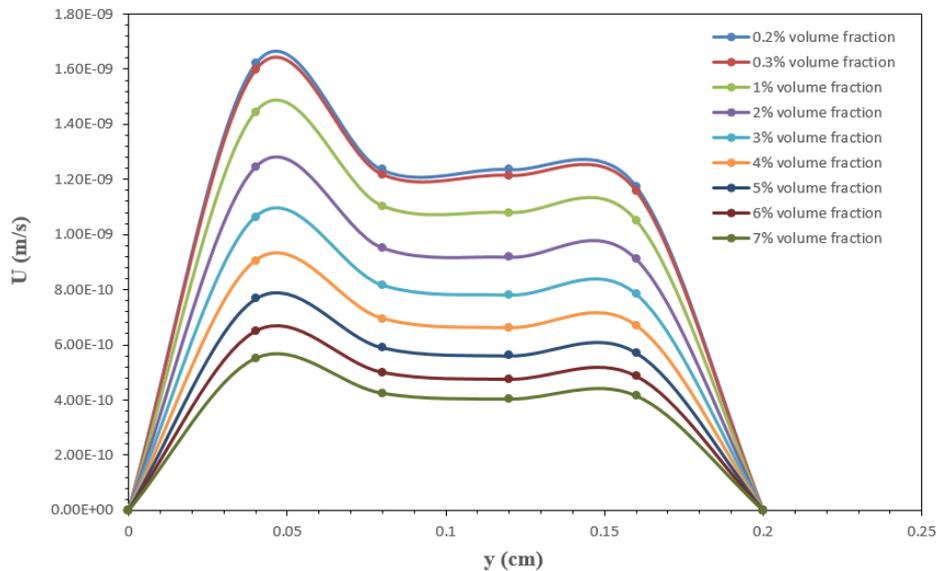

**Figure 8:** Velocity over a thickness of 2 mm cavity in different volume fractions for Water- Aluminum Oxide nanofluid

### 3.2.3. Ethylene Glycol-Al$_2$O$_3$ Nanofluid

The correlation between conductivity coefficients with volume fractions in four distances (the thicknesses that only conduction heat transfer takes place) is shown in Figure 9. With the addition of nanoparticles in ethylene glycol base fluid, thermal conductivity significantly increases. Besides, the thermal conductivity grows with volume fraction increment.

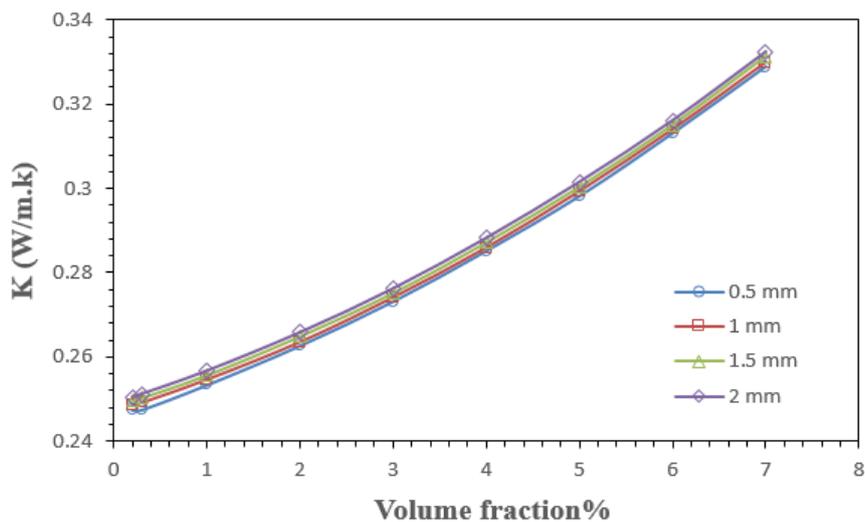

**Figure 9:** Changes in thermal conductivity of Ethylene Glycol- Aluminum Oxide nanofluid in various thicknesses of cavity

Owing to the high viscosity of ethylene glycol in comparison to water, nanofluid flow lines will be obtained in two different modes. In distance greater than 2 mm, convective heat transfer occurs and flow rotation arises. Two nanofluid flow lines in the water and ethylene glycol base fluid are plotted in Figure 10. Regarding the flow lines, it can be observed that in the water based nanofluid, two loops will be formed and in ethylene glycol based nanofluid one loop will be formed. The underlying reason for this phenomenon is the viscosity of the base fluid.

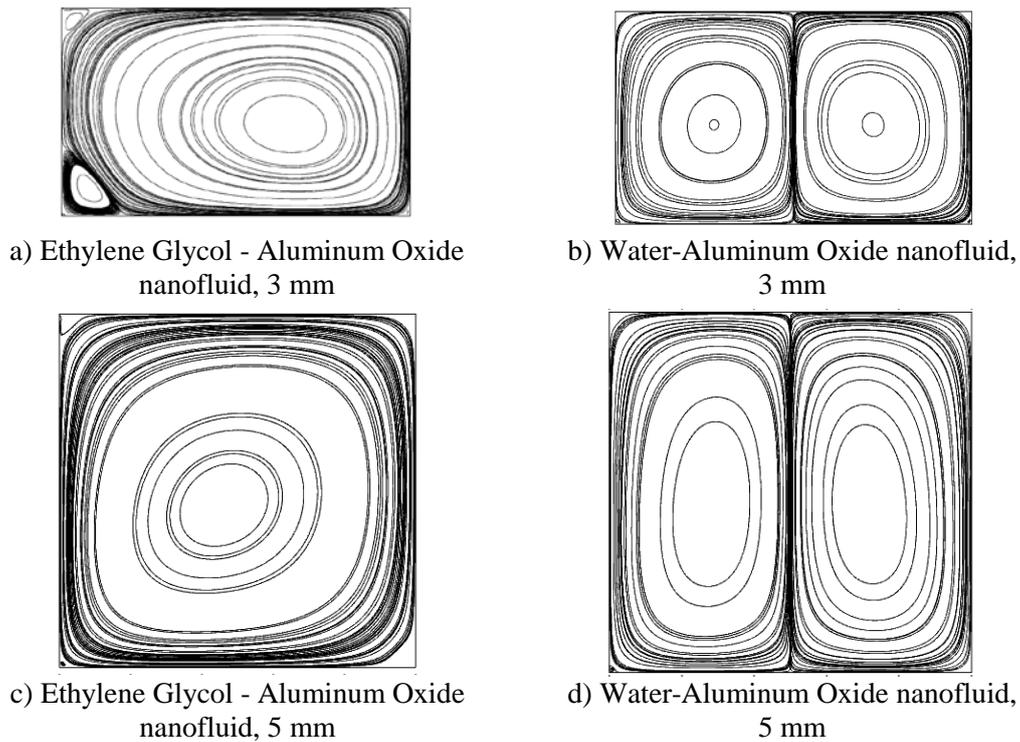

a) Ethylene Glycol - Aluminum Oxide nanofluid, 3 mm

b) Water-Aluminum Oxide nanofluid, 3 mm

c) Ethylene Glycol - Aluminum Oxide nanofluid, 5 mm

d) Water-Aluminum Oxide nanofluid, 5 mm

**Figure 10:** Comparison of flow lines of Water- Aluminum Oxide with Ethylene Glycol - Aluminum Oxide

### 3.2.4. Effect of Nanoparticle Size

In this section, four different Aluminum Oxide diameter values: 13, 28, 47 and 72 nm have been perused. Equation 10 has been applied to investigate the effects of geometry and calculate the thermal conductivity of nanofluid. Figures 11 and 12 show the size of nanoparticles effect in the water - aluminum oxide and ethylene glycol - aluminum oxide nanofluids. According to the results of the previous conditions, with volume fraction increment, nanoparticles thermal conductivity increases. Also, diameter of nanoparticles is inversely proportional to the thermal conductivity.

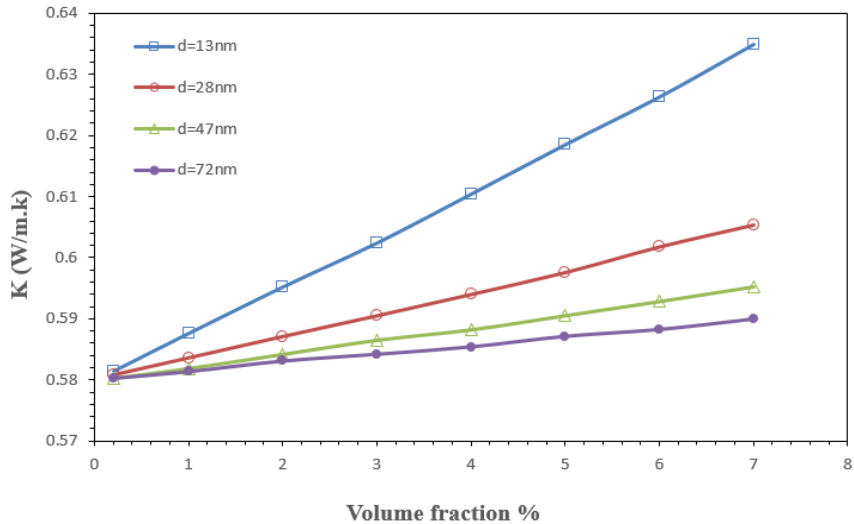

**Figure 11:** changes in the thermal conductivity with different diameters of nanoparticles and different volume fraction in Water -Aluminum Oxide nanofluid

According to Figures 11 and 12, the size of the diameter of 13 nm gives the best thermal conductivity. Moreover, slope of the graph is reduced with increasing diameter, which means that by increasing the diameter, volume fraction is less effective in increasing the conductivity.

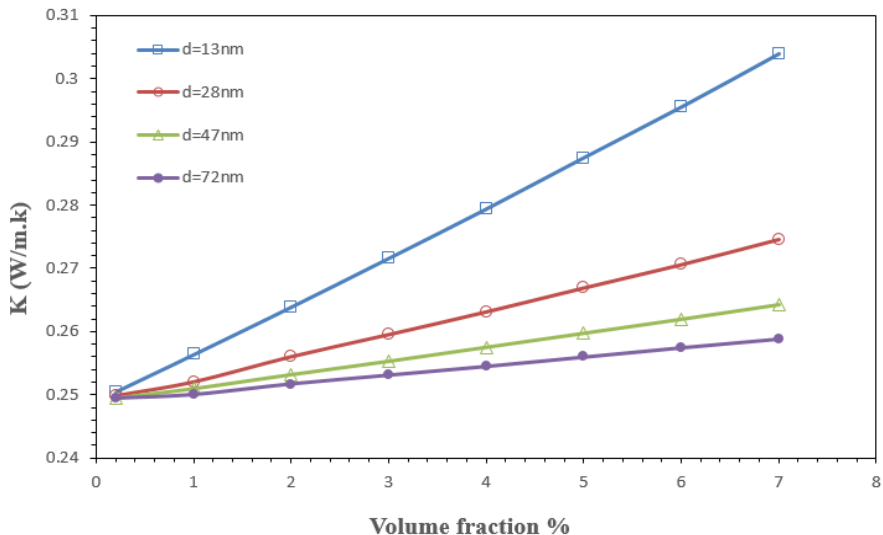

**Figure 12:** changes in the thermal conductivity with different diameters of nanoparticles and different volume fraction in Ethylene Glycol -Aluminum Oxide nanofluid

## 4. Conclusion

In the present work, heat conduction of nanofluids in a cylindrical cavity is studied by the parallel steady-state method. Three crucial factors' impacts on conductive heat transfer coefficient have been investigated which are the distance between two parallel discs, the base fluid and the volume fraction of nanoparticles. In the utilized method for modeling a cylindrical cavity, the foremost factor influencing the heat conduction is the distance between two parallel discs. Different distances between two discs were surveyed to determine the range of heat conduction. As a result, for a distance less than 2 mm and Ra<1000, the heat transfer by conduction occurs and for a distance more than them, the natural convection heat transfer takes place. By adding nanoparticles to the base fluid, thermal conductivity will increase. To optimize this method, increase in volume fraction will have a positive impact. Furthermore, it is construed that diameter of the nanoparticle is inversely proportional to the thermal conductivity of nanofluid.